\documentclass [twocolumn,amsmath,amssymb,prb,showpacs] {revtex4}

\usepackage{graphicx}

\begin{document}

\preprint{\empty}

\title{Diffusion and activation of $n$-type dopants in germanium\footnote{Copyright 2008 American Institute of Physics. This article may be downloaded for personal use only. Any other use requires prior permission of the author and the American Institute of Physics. The following article appeared in J.~Appl.~Phys. \textbf{104}, 023523 (2008) and may be found at http://link.aip.org/link/?JAP/104/023523.}}%

\author{Masahiro~Koike}
\email{m-koike@mail.rdc.toshiba.co.jp}%
\author{Yoshiki~Kamata}%
\author{Tsunehiro~Ino}%
\author{Daisuke~Hagishima}%
\author{Kosuke~Tatsumura}
\author{Masato~Koyama}%
\author{Akira~Nishiyama}%

\affiliation{
Corporate Research \& Development Center, Toshiba Corporation, 1 Komukai Toshiba-cho, Saiwai-ku, Kawasaki 212-8582, Japan
}%

\date{\today}

\begin{abstract}
\noindent The diffusion and activation of $n$-type impurities (P and As) implanted into $p$-type Ge(100) substrates were examined under various dose and annealing conditions. The secondary ion mass spectrometry profiles of chemical concentrations indicated the existence of a sufficiently high number of impurities with increasing implanted doses. However, spreading resistance probe profiles of electrical concentrations showed electrical concentration saturation in spite of increasing doses and indicated poor activation of As relative to P in Ge. The relationships between the chemical and electrical concentrations of P in Ge and Si were calculated, taking into account the effect of incomplete ionization. The results indicated that the activation of P was almost the same in Ge and Si. The activation ratios obtained experimentally were similar to the calculated values, implying insufficient degeneration of Ge. The profiles of P in Ge substrates with and without damage generated by Ge ion implantation were compared, and it was clarified that the damage that may compensate the activated $n$-type dopants has no relationship with the activation of P in Ge.
\end{abstract}


\pacs{66.30.J-, 61.72.Cc, 61.72.uf, 79.20.Rf, 82.80.Ms, 72.80.Cw}
%


\maketitle

\section{\label{sec:level1}INTRODUCTION}

The shrinkage of electrical devices has proceeded in order to achieve next-generation large-scale integrated circuits (LSIs). However, efforts to achieve continued downscaling in accordance with the road map \cite{ITRS} will encounter difficulties. Therefore, SiO$_{2}$ film, which is currently used as a gate insulator, is expected to be replaced with high dielectric constant (high-$k$) materials such as HfSiON (Ref. \onlinecite{koike2006}) and LaAlO (Ref. \onlinecite{suzuki2005}) to further reduce the electrical thickness without the reduction in physical thickness. Since one of the reasons for the use of Si substrates is that SiO$_{2}$ film of good quality can be grown by thermal oxidation on Si surfaces, other materials could be used as substrates instead of Si if high-$k$ materials were used instead of SiO$_{2}$.%

Germanium, which was the first semiconductor used at the dawn of semiconductor technology, is again being promoted as an alternative substrate to Si. One of the reasons for the replacement of Ge with Si was the difficulty in growing stable germanium oxide. This issue, however, becomes irrelevant when high-$k$ materials are employed instead of SiO$_{2}$. A characteristic that makes Ge attractive is the higher mobility in Ge than in Si. The mobilities of electrons and holes in Ge are 1900 and 3900~cm$^2$/Vs, respectively, whereas the corresponding mobilities in Si are 450 and 1500~cm$^2$/Vs, respectively. \cite{sze} Therefore, higher surface mobilities of electrons and holes are also expected in metal-insulator-semiconductor field effect transistors (MISFETs) with Ge substrates than in those with Si substrates. However, these higher mobilities have only been demonstrated in $p$-channel MISFETs. \cite{kamata2005, chui2002iedm} To the best of our knowledge, there have been no reports on high mobility in $n$-channel MISFETs with Ge substrates. This is mainly due to the high contact resistance at metal/$n^+$Ge junction, which prevents precise measurement of mobility. The high contact resistance is a result of the difficulty in obtaining $n^{+}$Ge layers with sufficiently high electrical concentrations (carrier concentrations), especially in the case of fabrication by ion implantation. \cite{chui2003apl, chui2005} In addition, it has been reported that $n$-type impurities such as P and As show concentration-dependent diffusion \cite{chui2003apl, matsumoto1978, vainoneon2000, vainoneon2001, bracht2006, bracht2007, brotzmann2008} that can be modeled by a dopant-vacancy-pair (AV)$^{-}$ mechanism. \cite{bracht2006, bracht2007, brotzmann2008} Further investigation is required in order to accurately control the diffusion for the production of the source/drain regions for future-generation LSIs. Thus, it is of scientific as well as technological interest to study why $n$-type impurities in Ge show such a low level of electrical activation and how various conditions, such as the implanted dose and annealing temperature, affect their diffusion.%

In this paper, the profiles of the chemical and electrical concentrations of $n$-type impurities in Ge under various conditions were investigated to clarify the diffusion mechanism and the electrical activation of $n$-type dopants in Ge. In addition, the effects of ion implantation damage on the activation were studied in detail. The profiles revealed that the electrical concentrations tend to become saturated in the case of heavy doses of impurity, whereas the chemical concentrations increase with implanted doses of impurity. The relationship between the chemical and electrical concentrations of P in Ge was calculated, and it was found that the activation ratio of P in Ge is comparable to that in Si because of the shallower impurity level of P and the lower conduction band density of state (DOS) in Ge. The activation levels of As in Ge were lower than those of P in Ge. The profiles of the chemical and electrical concentrations of P in Ge showed a concentration-dependent diffusion, which could be explained by the dopant-vacancy-pair (AV)$^{-}$ mechanism. The highest activation ratio data value did not exceed the incomplete ionization ratio value obtained by calculation, taking into account the single donor level, implying insufficient formation of the impurity band. In addition, it was confirmed that the difference in the activation ratios of P and As in Ge has no relationship with the damage resulting from ion implantation.%

\section{EXPERIMENT}
The substrates used in this study were 3 in. $p$-type Ge(100) wafers with Ga dopants of 2$\times$10$^{16}$~cm$^{-3}$. The surfaces of the wafers were first treated with 1\% HF solution to remove native oxide, then rinsed with purified water, and finally covered with thin SiO$_{2}$ layers ($\sim$4~nm) by reactive sputtering. The SiO$_{2}$ prevents outdiffusion of impurities. P or As ion was implanted as a $n$-type impurity into $p$-type Ge(100) substrates. The ion beams were tilted by 7$^{\circ }$ with respect to the normal to the surface. The implantation doses were changed from 2.0$\times 10^{14}$ to 5.0$\times 10^{15}$~cm$^{-2}$. The acceleration energies, 30 keV for As and 60 keV for P, were determined so as to achieve the same projected range value. For activation, the samples were annealed for 30 min at 400, 500, and 600 $\char'27\kern-.3em\hbox{C}$ in a N$_{2}$ ambient. Secondary ion mass spectrometry (SIMS) and the spreading resistance probe method (SRP) were employed to examine the depth profiles of the impurities. SIMS was used to estimate the profiles of the chemical concentrations of the dopants in Ge. The measurement was performed from the surface side of the specimens. SRP was used to examine the profiles of the electrical concentrations at room temperature. The profiles of the electrical concentrations were considered to be those of the activated dopants.%

\section{RESULTS and DISCUSSION}

\subsection{Diffusion of P in Ge: Implantation dose and annealing temperature dependence}\label{sec1}

\begin{figure}[t]
\includegraphics[width=8.5cm]{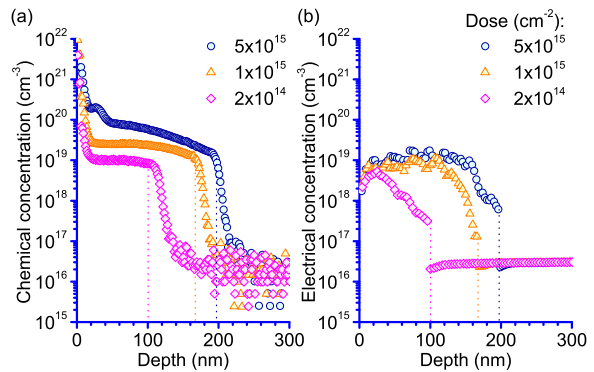}
\caption{\label{fig1} (Color online) Profiles of (a) chemical concentration and (b) electrical concentration of P in Ge substrates. The samples were annealed at 500$\char'27\kern-.3em\hbox{C}$ in a N$_2$ ambient for 30 min. The doses were 0.2, 1.0, and 5.0$\times 10^{15}$~cm$^{-2}$. Each broken line separates the profiles into two regions: the region where the concentration decreases slightly (left side) and the region where it decreases sharply (right side).}
\end{figure}

Figure \ref{fig1}(a) shows the dose dependence of the depth profiles of the chemical concentrations for P in Ge after annealing at 500 $\char'27\kern-.3em\hbox{C}$. The impurities diffuse into the Ge substrate and the concentrations increase with the increase in implantation dose. The total doses estimated from the profiles are identical with implanted doses. The concentrations around the surfaces reached at least 1$\times$10$^{19}$~cm$^{-3}$. Note that the profiles are separated into two regions according to the difference in behavior [the broken lines in Fig. \ref{fig1}(a) indicate the boundaries between the regions]; the concentration around the surface (the left side of the broken line) shows a slight decrease, whereas that within the substrate (the right side of the broken line) shows an abrupt drop. The concentrations at the boundaries are $\sim$1$\times$10$^{19}$~cm$^{-3}$. The behavior of the profiles in this study is the same as that observed in previous studies. \cite{chui2003apl, chui2005}%

A diffusion mechanism based on the dopant-vacancy-pair (AV)$^{-}$ model has previously been proposed for $n$-type dopants in Ge. \cite{bracht2006, bracht2007, brotzmann2008} The reaction of ionized P on substitutional sites with a singly positive charge P$_{S}^{+}$, and a vacancy with a doubly negative charge V$^{2-}$, generates singly negative dopant-vacancy pairs (PV)$^{-}$:
\begin{equation}
(\textrm{PV})^{-} \rightleftharpoons \textrm{P}_{S}^{+} + \textrm{V}^{2-}.
\label{reaction}
\end{equation}
Each of the concentrations of (PV)$^{-}$ and P$_{S}^{+}$ is represented by a diffusion-reaction equation and is connected through relationship (\ref{reaction}) (the law of mass action). The direct diffusion of P$_{S}^{+}$ is negligible and the dominant diffusion vehicle is (PV)$^{-}$, leading to the following single diffusion equation as a function of the concentration of P$_{S}^{+}$ and the effective diffusion coefficient: \cite{bracht2007}
\begin{equation}
\frac{\partial C_{\textrm{P}_{S}^{+}}}{\partial t} = \frac{\partial}{\partial x}
\left[ D_\textrm{eff}(n) \frac{\partial C_{\textrm{P}_{S}^{+}}}{\partial x} \right],
\label{diffeq}
\end{equation}
where
\begin{equation}
D_\textrm{eff}(n):=D^{\ast}(n_{i}) \left(\frac{ n}{n_{i}}\right)^2 .
\label{deff}
\end{equation}
$C_{\textrm{P}_{S}^{+}}(x,t)$ is the concentration of P$_{S}^{+}$, $n(x,t)$ is the electrical concentration, $n_{i}$ is the intrinsic carrier concentration dependent on temperature, $x$ is the depth from the surface, and $t$ is the diffusion time. $D^{\ast}(n_{i})$ is a diffusion coefficient associated with the intrinsic carrier concentration, while $D_\textrm{eff}(n)$ is an effective diffusion coefficient that has quadratic dependence on the carrier concentration. This dependence is due to the difference in the charge states between P$^{+}_{S}$ and (PV)$^{-}$, which has its origins in the indirect diffusion of P$^{+}_{S}$ through the formation of (PV)$^{-}$. \cite{bracht2006, bracht2007, brotzmann2008}

The shapes of the profiles in the results can be explained by referring to this model. In the region of low impurity concentration, $D_\textrm{eff}(n) = D^{\ast}(n_{i})$, i.e., the diffusion coefficient is constant. In the region of high impurity concentration, on the other hand, $D_\textrm{eff}(n)$ increases with the impurity concentration owing to the quadratic dependence on the carrier concentration.%

Using this mechanism, it is possible to examine the diffusion from the viewpoint of electrical concentration. Figure \ref{fig1}(b) shows the depth profiles of the electrical concentrations for P in Ge. The profiles are separated into two regions (the broken lines indicate the boundaries between the regions), similar to those for the chemical concentrations [Fig. \ref{fig1}(a)]. The substrate [right side of the boundary in Fig. \ref{fig1}(b)] includes a region in which the distribution of holes is uniform, originating from Ga dopants in Ge. On the other hand, a large number of electrons were observed around the surface [left side of the boundary in Fig. \ref{fig1}(b)], indicating the existence of $n$-type impurities, i.e., P. Note that the depth where the electrical concentrations abruptly decrease corresponds to the depth where the chemical concentrations also abruptly decrease. This is consistent with Eq. (\ref{deff}), which suggests that the diffusion decreases as the electrical concentrations decrease. The shapes of the profiles [Fig. \ref{fig1}(a)] can thus be explained by faster diffusion at higher concentrations than at lower concentrations.%

\begin{figure}[t]
\includegraphics[width=8.5cm]{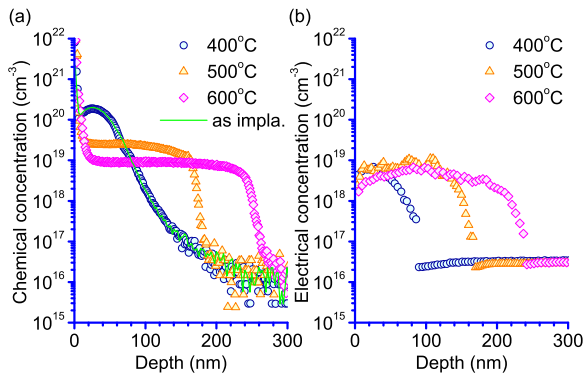}
\caption{\label{fig2} (Color online) Profiles of (a) chemical concentration and (b) electrical concentration of P in Ge substrates. The dose in each case was 1.0$\times 10^{15}$~cm$^{-2}$. The annealing temperatures were 400, 500, and 600 $\char'27\kern-.3em\hbox{C}$. The solid line in (a) indicates the profile for the as-implanted sample.}
\end{figure}

The temperature dependence of diffusion was also investigated. Figures \ref{fig2}(a) and \ref{fig2}(b) show the profiles of the chemical and electrical concentrations, respectively, of P in Ge substrates that were annealed at 400, 500, and 600$\char'27\kern-.3em\hbox{C}$. The profile of the chemical concentration for a sample before annealing is also shown in Fig. \ref{fig2}(a) as a solid line. The tail in the profile is due to ion channeling. The profile is nearly the same as that for the sample annealed at 400$\char'27\kern-.3em\hbox{C}$, indicating that no diffusion occurred at this temperature. 

With the increase of annealing temperatures, impurities spread deep into Ge substrates [Fig. \ref{fig2}(a)]. At the time, electrically activated impurities diffuse into the Ge substrates, keeping the activation levels almost constant [Fig. \ref{fig2}(b)]. Interestingly, the total number of activated dopants increases with temperature, even though the dose is not changed. It thus appears that electrically activated impurities are generated as the temperature is increased. As shown in Fig. \ref{fig2}(a), a large number of inactive impurities, beyond the saturation levels for activated impurities, exist around the surface. This suggests that inactive impurities change into activated impurities to maintain the saturation level during diffusion.%

\subsection{Relationship between the chemical and electrical concentrations}\label{sec2}

\begin{figure}[t!]
\includegraphics[width=8.5cm]{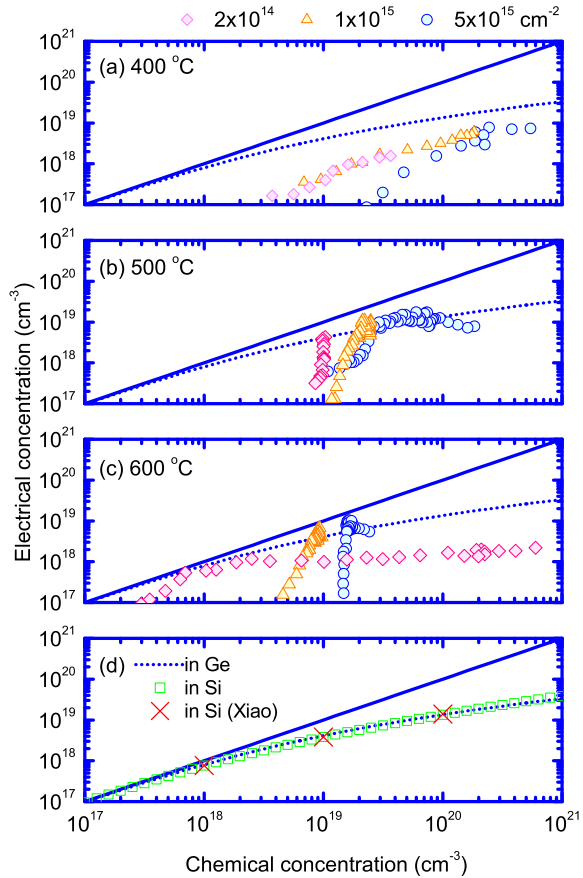}
\caption{\label{fig3} (Color online) Relationship between the chemical concentration and the electrical concentration of P in Ge substrates. The concentrations were measured at the same depth using SIMS and SRP. The samples were annealed at (a) 400, (b) 500, and (c) 600$\char'27\kern-.3em\hbox{C}$. The solid lines correspond to an activation ratio of 100\%. The dotted lines indicate the calculated electrical concentrations of P in Ge as functions of the chemical concentrations at 300~K, taking the effects of incomplete ionization into consideration. The calculated results obtained for P in Si by the authors and by Xiao \textit{et al}. (Ref. \onlinecite{xiao1999}), whose results are similar, are shown in (d).}
\end{figure}

Figure \ref{fig3} shows the relationship between the chemical and electrical concentrations of P in Ge at various temperatures. The concentrations were obtained using SIMS and SRP data. The chemical concentrations at the depth where SRP was performed were interpolated from the SIMS data to enable comparison of the measurement results of the two methods at the same depth. The solid line indicates the result for fully activated P in Ge, while the dotted line indicates the result taking into account incomplete ionization,\cite{altermatt2006, altermatt2006-2, xiao1999, watanabe2001} the details of which will be shown later. The data for higher doses tends to distribute in the region of higher chemical concentration. As the temperature becomes higher, the chemical concentrations tend to be lower, owing to the faster diffusion of dopants. On the other hand, higher temperatures can create higher electrical concentrations, i.e., they can lead to higher activation ratios. A temperature of 500$\char'27\kern-.3em\hbox{C}$ or more is required to activate impurities; however, even at such high temperatures, the activation ratio does not reach 100\%. 

Let us consider why dopants have a much lower electrical activation in Ge than in Si. One of the possibilities is the low DOS; the effective DOS in the conduction band of Ge (1.04$\times 10^{19}$~cm$^{-3}$) is lower than that in the conduction band of Si (2.8$\times 10^{19}$~cm$^{-3}$).\cite{sze} It is thought that the lower DOS enhances incomplete ionization \cite{altermatt2006, altermatt2006-2, xiao1999, watanabe2001} in Ge, in contrast to the case of Si, because electrons need excitation to move from the donor level to higher energy levels of the conduction band. The relationships between the chemical and electrical concentrations of P in Ge and in Si were calculated to estimate how a low DOS affects the ionization ratio (the activation ratio). In the calculation, the Fermi level that satisfied the charge neutrality was searched.
\begin{equation}
n = N_{D} - n_{D},
\label{cn}
\end{equation}
\begin{equation}
n = \int _{E_{C}}^{\infty} \frac{D_{C}(E)}{e^{(E-E_{F})/k_{B}T}+1} dE,
\label{n}
\end{equation}
\begin{equation}
n_{D} = \frac{N_{D}}{\frac{1}{2}e^{(E_{D}-E_{F})/k_{B}T}+1}.
\label{nd}
\end{equation}
Here, $n$ is the electron concentration in the conduction band, $D_{C}(E)$ is the conduction band DOS, $N_{D}$ is the impurity concentration, and $n_{\textrm D}$ is the electron concentration in the donor level, which is assumed to be a single level. As shown by the dotted curves in Figs. \ref{fig3}(a)-\ref{fig3}(c), the activation ratio decreases with the chemical concentration. Coincidentally, this curve is almost the same as that for Si [Fig. \ref{fig3}(d)], which is similar to the curve for Si obtained by Xiao \textit{et al.} \cite{xiao1999} This is because the donor level of P in Ge ($E_{C}-E_{D}$=12~meV) is shallower than that in Si (45~meV). \cite{sze} The ionization ratio tends to increase as the impurity level becomes shallower and as the DOS increases. The effect of the shallower donor level and that of the lower DOS compensate each other in this case. Thus, the ionization ratio of P in Ge can be regarded as almost the same as that in Si. In the case of Si, previous reports have indicated that the ionization ratio decreases with the doping concentration at low concentrations ($N_{D}<10^{18}$~cm$^{-3}$) owing to the incomplete ionization, and then starts to increase at high concentrations ($N_{D}>10^{18}$~cm$^{-3}$) owing to the formation of impurity bands \cite{altermatt2006, altermatt2006-2} in the band gap of Si. It is believed that Mott (metal-insulator) transition \cite{kittel, roessler, altermatt2006, altermatt2006-2} and the merging of the conduction band and impurity band \cite{altermatt2006, altermatt2006-2} are a possible mechanism for the metallic (degenerate semiconductor) properties in heavily doped Si.%

The experimental data values, especially those at 500 and 600$\char'27\kern-.3em\hbox{C}$, show a tendency to distribute along on the dotted curve (Fig. \ref{fig3}). It should be emphasized here that the dotted curve indicates the upper limit of the electrical concentration at a chemical concentration, calculated on the basis of the incomplete ionization effect in a single donor level. To exceed this curve, dopants need to form an impurity band as well as to entirely substitute in Ge. By analogy to the case of Si and in light of the shallower donor level and lower critical concentration of the Mott transition (2.57$\times 10^{17}$~cm$^{-3}$ for Ge, 3.5$\times 10^{18}$~cm$^{-3}$ for Si),\cite{kittel} the impurity band in Ge should be formed to reach the activation ratio of 100\% in the region of high concentration ($>10^{18}$~cm$^{-3}$). It is therefore speculated that the poor activation of impurities in Ge is a result of insufficient formation of the impurity band in the energy band gap, which may be caused by various complexes, such as dopant-vacancy pairs and/or clustering of impurities. A previous report has suggested the possible presence of P clusters in Ge at high concentrations of P.\cite{satta2006} When clusters are formed, they can be of two possible types. The first type is an electrically inactive cluster. In this case, of course, the cluster does not contribute to the formation of the impurity band and the number of impurities interacting with each other decreases or the distance between the impurities increases, causing poor activation. The second type is an electrically active cluster. Even in this case, the activation ratio would decline. Since impurities in clusters are locally aggregated, they would not interact significantly with nearby impurities or other clusters, leading to insufficient impurity band formation. Our experiments showed that the electrical concentration for the same chemical concentration tended to increase with the annealing temperature. It is therefore thought that clusters and/or some complexes are decomposed into impurities at higher temperature to form an impurity band, leading to higher activation ratios.%

\subsection{Effects of impurity differences and ion implantation damage on profiles}\label{sec3}

\begin{figure}[t!]
\includegraphics[width=8.5cm]{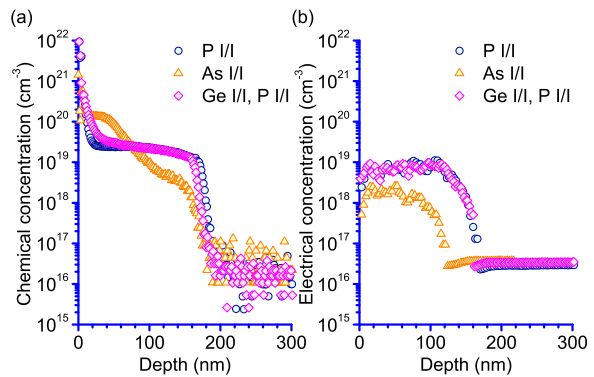}
\caption{\label{fig4} (Color online) Depth profiles of (a) chemical concentrations and (b) electrical concentrations of impurities in Ge substrates. Ge I/I, P I/I refers to the profiles of P in Ge substrates in which Ge was implanted first, followed by P. P I/I and As I/I, respectively, indicate the profiles of P and As implanted in Ge substrates in which Ge was not implanted. The annealing temperature for all samples was 500$\char'27\kern-.3em\hbox{C}$. The implanted dose in each case was 1$\times 10^{15}$~cm$^{-2}$.}
\end{figure}

Next, the effects of impurity differences and ion implantation damage on the profiles are examined. Figures \ref{fig4}(a) and \ref{fig4}(b) show the profiles of the chemical and electrical concentrations, respectively, of the impurities (P and As) in Ge. ``P I/I'' and ``As I/I,'' respectively, indicate the profiles of P and As implanted in Ge substrates. ``Ge I/I, P I/I'' refers to the profiles of P in Ge substrate where Ge was implanted first, followed by P. The annealing temperature was 500$\char'27\kern-.3em\hbox{C}$. The implanted dose was 1$\times 10^{15}$~cm$^{-2}$. It was found that the profiles of P in Ge with and without implanted Ge were almost the same, for both the chemical concentration and the electrical concentration. On the other hand, the profiles of P were different from those of As. The chemical concentrations of P and As were $\sim$2$\times$10$^{19}$ and $\sim$2$\times$10$^{20}$~cm$^{-3}$, respectively, whereas their electrical concentrations were $\sim$1$\times$10$^{19}$ and $\sim$2$\times$10$^{18}$~cm$^{-3}$, respectively.%

The differences between the profiles of P and As [Figs. \ref{fig4}(a) and \ref{fig4}(b)] and why the activation level of As is lower than that of P in Ge are now examined. Some studies have discussed these issues. \cite{vainoneon2000, vainoneon2001, chui2003apl,chui2005, dilliway2006} According to them, one of the possible explanations for the differences is the solid solubility. Chui \textit{et al.} \cite{chui2003apl,chui2005} have reported that the electrical concentrations of $n$-type impurities such as P and As implanted in Ge are saturated. A similar tendency was observed in our results. In their work, the saturation levels of P and As reached $\sim$5$\times$10$^{19}$ and $\sim$3.5$\times$10$^{19}$~cm$^{-3}$, respectively. They concluded that this difference was caused by solid solubility differences. Our understanding is that the solid solubility is generally related to the atomic radius, and larger atoms are substituted less in the Ge crystal. Therefore, As, due to its larger size, has a lower solid solubility than P. We think that the excess As impurities, which cannot be substituted in the Ge lattice as a result of their low solid solubility, form various complexes and/or clusters, leading to the poor activation of As. On the other hand, we also think that the formation of acceptorlike defects generated by ion implantation may contribute to the poor activation of As. Defects generated in Ge tend to form acceptorlike defects.\cite{trumbore1960} Such defects may compensate the donor levels of the $n$-type impurities, and therefore reduce the concentrations of electrons. Since the amount of damage generated in Ge tends to increase with the mass of the implanted atoms,\cite{satta2006} it is possible that the implantation of As causes more such defects in the Ge substrate than the implantation of P; however, this has not yet been clarified. This possibility can be clarified from the Ge I/I, P I/I profiles in Figs. \ref{fig4}(a) and \ref{fig4}(b). Since Ge has almost the same mass as As, the damage by implantation of Ge should be almost the same as that of As. If the damage by implantation causes poor activation of As relative to P, the electrical concentrations in the Ge substrates where both P and Ge were implanted should be less than that in the Ge substrates where only P was implanted. As shown in Figs. \ref{fig4}(a) and \ref{fig4}(b), however, there were no differences in the chemical or electrical concentrations of P in the two cases, and the concentration of P in the substrate was higher than that of As. Thus, the difference between the activation levels of P and As has no relationship with the damage caused by ion implantation.%

\section{SUMMARY}

The effects of ion implantation on the chemical and electrical concentrations of $n$-type dopants (As and P) in Ge substrates were investigated. Whereas the chemical concentrations around the surface increased with dose, the electrical concentrations tended to become saturated. The saturation level of As was lower than that of P. The behavior of the chemical and electrical concentrations could be explained based on the quadratic dependence of the diffusion coefficient on the carrier concentration, which has its origins in the dopant-vacancy-pair mechanism. The relationships between the chemical and electrical concentrations of P in Ge and in Si were calculated. The results indicated that the ionization ratio of P in Ge was comparable to that of P in Si owing to the effects of the shallower donor level in Ge relative to Si and the lower conduction band DOS in Ge relative to Si compensating each other. In addition, the low electrical activation of P in Ge was shown to be the result of insufficient formation of the impurity band required to reach the ionization ratio of 100\%. The profiles of impurities in Ge substrates with the same damage by ion implantation were compared, and it was clarified that the damage by ion implantation is not related to the difference in saturation levels of As and P in Ge.%

\section*{ACKNOWLEDGMENTS}

The authors would like to thank Y. Nakabayashi of Toshiba Corporation for his helpful comments and suggestions.%


\end{document}